\documentclass[pra,aps,twocolumn,epsfig,nofootinbib,superscriptaddress]{revtex4}

\usepackage{amsmath}
\usepackage{amssymb}
\usepackage[dvips]{graphicx}%\usepackage{graphicx}% Include figure files
\usepackage{dcolumn}% Align table columns on decimal point
\usepackage{bm}% bold math
\usepackage[colorlinks=false,dvipdfm]{hyperref} % citecolor=blue 文件内部引用
\usepackage{setspace}
\usepackage{amsfonts}
\usepackage{rotating}
\usepackage{makeidx}
\usepackage{amsthm}

\begin{document}
%\begin{CJK*}{GB}{}

\title{Perfect teleportation with a partially entangled quantum channel\footnote{Phys. Rev. A 106, 032430 (2022)}}

\author{Xiang Chen}
\affiliation{Department of Physics, School of Science, Tianjin University, Tianjin 300072, China}

\author{Yao Shen}
\affiliation{School of Criminal Investigation, People's Public Security University of China, Beijing 100038, China}

\author{Fu-Lin Zhang}
\email[Corresponding author: ]{flzhang@tju.edu.cn}
\affiliation{Department of Physics, School of Science, Tianjin University, Tianjin 300072, China}

\date{\today}

\begin{abstract}
Quantum teleportation provides a way to transfer unknown quantum states from one system to another via an
entangled state as a quantum channel without physical transmission of the object itself. The entangled channel,
measurement performed by the sender (Alice), and classical information sent to the receiver (Bob) are three
key ingredients in the procedure, which need to cooperate with each other. To study the relationship among
the three parts, we propose a scheme for perfect teleportation of a qubit through a high-dimensional quantum
channel in a pure state with two equal largest Schmidt coefficients. The scheme requires less entanglement of
 Alice's measurement but more classical bits than the original scheme via a Bell state. The two quantities increase
with the entanglement of the quantum channel when its dimension is fixed and thereby can be regard as Alice's
necessary capabilities to use the quantum channel. And the two capabilities appear complementary to each other.
\end{abstract}

 %\pacs{03.67.Ac, 03.65.Ta, 03.67.Bg, 03.67.Hk}

%\keywords{}

\maketitle
%\end{CJK*}

\section{Introduction}
Significant differences between the quantum  and classical worlds are revealed by several quantum information processes without classical counterparts \cite{Book} .
One of these processes is quantum teleportation \cite{bennett1993teleporting},
in which Alice (the sender) can transfer unknown quantum states from her system to Bob  (the receiver), without physical transmission of the object itself.
In the simplest and original form, the task of Alice is to teleport a qubit state to Bob.
They share a two-qubit Bell state as the quantum channel in advance.
Alice makes a joint measurement on the state to be teleported and her qubit from the Bell state,
projecting them onto one of the four Bell states.
After Alice informs him of the outcome through a classical channel,
Bob can perform appropriate unitary operations on his qubit to perfectly rebuild the state to be teleported.

The teleportation protocol has been  extended in many branches, including  probabilistic teleportation through a partially entangled pure state \cite{1999Probabilistic,Banaszek2000Optimal,Roa2003Optimal,2003Optimal,PRA2015RoaTele},
controlled teleportation involving  a third party as a controller \cite{Ctele,PhysRevA.90.052305},
teleportation in high dimensions \cite{HUANG2020Quantum,2019Quantum,2020Experimental}, and so on.
These schemes play key roles in various contexts in quantum communication, including in quantum repeaters, quantum
networks, and cryptographic conferences \cite{repeaters,PhysRevA.54.2651,PhysRevA.57.822,networks,QNet}.
What these versions of  teleportation have in common is that
the entanglement of quantum channels is destroyed by Alice's measurement and thereby is the cost of accomplishing the task.
Consequently,  teleportation serves as an important example for the quantum information processes, in which  entanglement plays the role of a key resource \cite{brunner2005entanglement,RevModPhys.81.865}.

The entangled quantum channels,
Alice's joint measurement, and the classical information sent to Bob should be regarded as three key ingredients in the procedure, which need to cooperate with each other.
Measurement of entangled states, as well as their generation, is a technical challenge in laboratories \cite{2019Quantum,2020Experimental}
which theoretically can be implemented by using the inverse process of entanglement preparation and local measurements \cite{Book}.
The primary objective of this work is to study the relationships among the three key ingredients in the procedures for teleportation.
We focus on the perfect teleportation (with $100\%$ success probability and fidelity) of a qubit, the smallest unit of quantum information.

Our first step is to propose a general scheme for perfect teleportation of  a qubit by using
partially entangled two-qudit (a $d$-dimensional quantum system with $d\geq3$) states,
which  form continuous regions in the spaces of entanglement invariants  \cite{RevModPhys.81.865,Fei,RUI2011Alternative}.
In most of the protocols for perfect  teleportation
\cite{bennett1993teleporting,HUANG2020Quantum,2019Quantum,2020Experimental},
quantum channels are limited to maximally entangled states, locating at vertices of the areas of entanglement invariants.
%,
To the best of our knowledge,
Gour \cite{PRA2004} proposed the only protocol for perfect teleportation via partially entangled two-qudit states in the literature\footnote{\textbf{Note added after publication}.--Recently, the authors became aware of a related teleportation scheme presented in [\emph{Quantum Bio-Informatics II: From Quantum Information to Bio-Informatics}, 19-29, (2009)] and  [Reports on Mathematical Physics  69, 57 (2012)].}.
However,  in Gour's protocol, the classical bits are fixed to be the logarithm of the total dimension of the two subsystems in Alice's hands,
instead of relying on the entanglement of quantum channel.
Here, we propose a general scheme for perfect teleportation of  a qubit with a lower classical communication cost
by using two-qudit pure states in which the two largest Schmidt coefficients are equal.
These quantum channels are  coherent superpositions of a set of  maximally entangled states in subspaces
which form a $(d-2)$-dimensional polyhedron in the space of entanglement invariants  \cite{RevModPhys.81.865,Fei,RUI2011Alternative}.

To measure Alice's effort in her joint measurement, we define  the entanglement of measurement as the average entanglement of its basis.
In general, our scheme requires less entanglement of  Alice's measurement but more classical bits than the ones with Bell states.
The two quantities can be regard as Alice's necessary capabilities to use the quantum channel,
which  increase with its entanglement when the dimension is fixed.
And the two abilities are complementary to each other, as the price to decrease the entanglement in the measurement is to send more classical bits to Bob.

In the next section, we explain  our   protocol by using the example of a three-level entangled quantum channel.
The general scheme is presented in Sec. \ref{qudit}.
The relationship among the entangled channel,
the entanglement of Alice's measurement, and the classical information in the task are studied in Sec. \ref{entanglement}.
In Sec. \ref{disscuss},
we give an intuitive understanding of the relationship
and show the robustness of our protocol under an inhomogeneous phase noise.
Finally, Sec. \ref{summ} presents a summary.

 \section{Two-qutrit quantum channel}	\label{qutrit}
%\textit{Two-qutrit quantum channel.--}
Let us first explain  our basic idea to design the protocol by using the example of a two-qutrit (three-level systems) entangled quantum channel.
Suppose Alice wishes to teleport to Bob the qubit state
\begin{equation}
\left\vert \phi\right\rangle _{1}=\alpha\left\vert 0\right\rangle _{1}+\beta\left\vert 1\right\rangle _{1},
\end{equation}
with $|\alpha|^2+|\beta|^2=1$,  and they share a two-qutrit entangled quantum channel
\begin{equation}
\left\vert \Phi\right\rangle _{23}=a_{0}\left\vert 00\right\rangle _{23}+a_{1}\left\vert 11\right\rangle _{23}+a_{2}\left\vert 22\right\rangle _{23},
\end{equation}
with $\sum_{j=0}^2|a_{j}|^2=1$.
Without loss of generality, one can assume the Schmidt coefficients $a_{j=0,1,2}$ are real numbers and  $0\leq a_0 \leq a_1\leq a_2 $.
Here, we set $a_1 = a_2$.
Then, the entanglement entropy \cite{Ent1996,Wootters98} of  $\left\vert \Phi\right\rangle _{23}$ is larger than $1$ and
is a monotone increasing function of  $a_0$.
And in the space of entanglement invariants \cite{RUI2011Alternative},  the points of $\left\vert \Phi\right\rangle _{23}$ form one of the three edges of the region of arbitrary two-qutrit pure states.
The two extreme cases,
\begin{subequations}\label{max3}
\begin{align}
\!\!\!&\!\!\!\left\vert \Phi\right\rangle _{23}^{(a)}\!=\!\frac{1}{\sqrt{3}}( \left\vert 00\right\rangle _{23}+ \left\vert 11\right\rangle _{23}+ \left\vert 22\right\rangle _{23}),\label{max3a}   \\
\!\!\!&\!\!\!\left\vert \Phi\right\rangle _{23}^{(b)}\!=\!\frac{1}{\sqrt{2}}(\left\vert 11\right\rangle _{23}+ \left\vert 22\right\rangle _{23}), \label{max3b}
\end{align}
\end{subequations}
locate at two end points of the edge.
The former is the maximally entangled two-qutrit state, and the latter is equivalent to the two-qubit Bell states.
Based on these entanglement properties of  $\left\vert \Phi\right\rangle _{23}$, we expect it can be adopted as the quantum channel to  teleport
 the qubit state $\left\vert \phi\right\rangle _{1}$ perfectly.

The most crucial step for
%The key step of
the teleportation is Alice's joint measurement on her qubit 1 and qutrit 2,
which projects Bob's qutrit 3 into a state dependent on Alice's measurement result and the state $\left\vert \phi\right\rangle _{1}$.
After Alice informs Bob of her measurement result through a classical channel, Bob performs a corresponding operation on his system to recover $\left\vert \phi\right\rangle _{3}=\alpha\left\vert 0\right\rangle _{3}+\beta\left\vert 1\right\rangle _{3}$.
To teleport the state perfectly (with a fidelity of $1$ and success probability of $1$), two conditions should be satisfied as follows:
(i)   Alice's measurement is a projective one;
(ii) the collapsed states of particle 3 are of the form $\alpha\left\vert \tilde0\right\rangle+\beta\left\vert \tilde1 \right\rangle$, with $\left\vert \tilde0\right\rangle$ and $\left\vert \tilde1\right\rangle$ being two orthogonal states independent of $\alpha$ and $\beta$.
The contrasting cases are the two schemes for probabilistic teleportation \cite{1999Probabilistic,Banaszek2000Optimal,Roa2003Optimal,2003Optimal,PRA2015RoaTele}
in which Alice unambiguously discriminates nonorthogonal states
or
Bob performs an extracting quantum state process.

To construct the basis of Alice's measurement, we write the total tripartite state as
\begin{align}
\left\vert \Psi\right\rangle _{123}  &  =\left\vert \phi\right\rangle_{1}\left\vert \Phi\right\rangle _{23}\nonumber\\
&  =\bigr[ a_{0}\left\vert 00\right\rangle\alpha\left\vert 0\right\rangle+a_{1}\left\vert 01\right\rangle \alpha\left\vert 1\right\rangle+a_{1}\left\vert 02\right\rangle \alpha\left\vert 2\right\rangle\\
&\ \ \   +a_{0}\left\vert 10\right\rangle\beta\left\vert 0\right\rangle+a_{1}\left\vert 11\right\rangle\beta\left\vert 1\right\rangle
+a_{1}\left\vert 12\right\rangle\beta\left\vert 2\right\rangle\bigr]_{123} \nonumber \ \ \
\end{align}
For the two extreme cases in Eqs. (\ref{max3}), one can easily find a \textit{translation strategy} from the above form in which
Alice operates a measurement with the eigenstates
\begin{align}\label{psia}
\left\vert \psi^{(a)}_{0\pm}\right\rangle_{12}  &  =\frac{1}{\sqrt{2}}(\left\vert00\right\rangle \pm\left\vert 11\right\rangle )_{12},\nonumber\\
\left\vert \psi^{(a)}_{1\pm}\right\rangle_{12}  &  =\frac{1}{\sqrt{2}}(\left\vert01\right\rangle \pm\left\vert 12\right\rangle )_{12},\\
\left\vert \psi^{(a)}_{2\pm}\right\rangle_{12}  &  =\frac{1}{\sqrt{2}}(\left\vert02\right\rangle \pm\left\vert 10\right\rangle )_{12},\nonumber
\end{align}
for case (\ref{max3a}), while she measures the entangled states
 \begin{align}\label{psib}
\left\vert \psi^{(b)}_{1\pm}\right\rangle_{12}  &  =\frac{1}{\sqrt{2}}(\left\vert01\right\rangle \pm\left\vert 12\right\rangle )_{12},\\
\left\vert \psi^{(b)}_{2\pm}\right\rangle_{12}  &  =\frac{1}{\sqrt{2}}(\left\vert02\right\rangle \pm\left\vert 11\right\rangle )_{12},\nonumber
\end{align}
for case (\ref{max3b}).
The former was studied in a recent work \cite{HUANG2020Quantum},
 and the latter is  precisely the original scheme \cite{bennett1993teleporting} with the simple substitutions $\left\vert0\right\rangle \rightarrow \left\vert1\right\rangle $ and $\left\vert1\right\rangle \rightarrow \left\vert2\right\rangle $ in subsystem 2.
 In fact, to be complete, the  basis for (\ref{max3b}) also contains two \textit{vanished} states, $\left\vert00\right\rangle $ and $\left\vert10\right\rangle$, whose probabilities are zero in the measurement.

One may expect  that the orthogonal basis of  Alice's measurement, for an arbitrary $\left\vert \Phi\right\rangle _{23}$ with $a_1=a_2$,
is intermediate states between the two sets of extreme basis.
A natural idea is to  reduce the proportions of $\left\vert00\right\rangle $ and $\left\vert10\right\rangle $ in   $\vert \psi^{(a)}_{0\pm}\rangle_{12} $ and $\vert \psi^{(a)}_{2\pm}\rangle_{12} $, while inserting the terms   $\left\vert02\right\rangle $ and $\left\vert11\right\rangle $, respectively, as their replacement.
However, this breaks the orthogonality in condition (i), mainly because it is a continuous transition from the four states to two.
Thus, we suspect that the two postmeasured states of qutrit 3 corresponding to $\vert \psi^{(b)}_{2\pm}\rangle_{12} $ can be  collapsed to four orthogonal
states with the aid of  the two \textit{vanished} vectors, $\left\vert00\right\rangle $ and $\left\vert10\right\rangle $.
One option is that we can replace $\vert \psi^{(b)}_{2\pm} \rangle_{12} $ by $ \vert \psi^{(b)}_{2+} \rangle_{12} \pm  \vert 10 \rangle_{12} $ and  $ \vert 00 \rangle_{12} \mp  \vert \psi^{(b)}_{2-} \rangle_{12} $ (which are un-normalized).
These can be connected smoothly  with $\vert \psi^{(a)}_{j\pm}\rangle_{12}$ by the six orthogonal bases
\begin{align}\label{psi3}
\left\vert \psi_{0\pm}\right\rangle _{12} &  =\frac{1}{\sqrt{2}}[\left\vert00\right\rangle \pm(c\left\vert 11\right\rangle -s\left\vert 02\right\rangle)]_{12},\nonumber\\
\left\vert \psi_{1\pm}\right\rangle _{12} &  =\frac{1}{\sqrt{2}}[\left\vert01\right\rangle \pm\left\vert 12\right\rangle ]_{12},\\
\left\vert \psi_{2\pm}\right\rangle _{12} &  =\frac{1}{\sqrt{2}}[(c\left\vert02\right\rangle +s\left\vert 11\right\rangle )\pm\left\vert 10\right\rangle ]_{12},\nonumber
\end{align}
with the two real numbers satisfying $c^2+s^2=1$.

Now, we show that the teleportation can be accomplished perfectly by performing a joint measurement   in the above basis (\ref{psi3}) with an appropriate pair of $c$ and $s$.
The (un-normalized) collapsed states of qutrit 3, corresponding to Alice's measurement results $\vert \psi_{j\pm} \rangle _{12} $, can be derived by
$
\vert \phi_{j\pm}\rangle _{3}=_{12}\!\!\langle \psi_{j\pm}\vert \Psi\rangle_{123},
$
which are
\begin{align}\label{psiipm}
\left\vert \phi_{0\pm}\right\rangle _{3} &  =\frac{1}{\sqrt{2}}[\alpha(a_{0}\left\vert 0\right\rangle _{3}\mp a_{1}s\left\vert
2\right\rangle _{3})\pm\beta a_{1}c\left\vert 1\right\rangle _{3}],\nonumber\\
\left\vert \phi_{1\pm}\right\rangle _{3} &  =\frac{1}{\sqrt{2}}[\alpha a_{1}\left\vert 1\right\rangle _{3}\pm\beta a_{1}\left\vert 2\right\rangle_{3}],\\
\left\vert \phi_{2\pm}\right\rangle _{3} &  =\frac{1}{\sqrt{2}}[\alpha a_{1}c\left\vert 2\right\rangle _{3}+\beta(a_{1}s\left\vert 1\right\rangle_{3}\pm a_{0}\left\vert 0\right\rangle _{3})].\nonumber
\end{align}
It is easy to find that they fulfill condition (ii) when
\begin{align}\label{cs}
\textstyle{
c= \sqrt{\frac{1}{2}\left(1+\frac{a_{0}^{2}}{a_{1}^{2}}\right)}\ \ \  \mathrm{,} \ \ \ s  = \sqrt{\frac{1}{2}\left(1-\frac{a_{0}^{2}}{a_{1}^{2}}\right)}.
}
\end{align}
Then,  the probabilities of Alice's outcome are given by the overlap $P_{j\pm}= _{3}\! \langle \phi_{j\pm}  \vert \phi_{j\pm}\rangle _{3}$ as
\begin{equation}
P_{0\pm}=P_{2\pm}=\frac{1}{4}(a_{0}^{2}+a_{1}^{2})\ \ \  \mathrm{,} \ \ \  P_{1\pm}=\frac{1}{2}a_{1}^{2}.
\end{equation}
Here, we omit Bob's unitary operators, to transform the state on his end to $|\phi\rangle_3$, which can be directly constructed by using the forms of $\vert \phi_{j\pm}\rangle _{3}$.

  \section{General protocol}	\label{qudit}
%\textit{General protocol.--}
Now we turn to the general protocol for teleporting a qubit through the following partially entangled two-qudit state as the quantum
channel:
\begin{equation}
\left\vert \Phi\right\rangle _{23}=
\sum_{i=0}^{n}
a_{i}\left\vert i\right\rangle _{2} \left\vert i \right\rangle _{3}
\end{equation}
where $n=d-1 =2,3,...$ and
the real Schmidt coefficients
%\begin{equation}
$ 0\leq a_{0}\leq a_{1}\leq a_{2}\leq ... \leq a_{n-1}=a_{n}$ and $\sum _{i=0}^{n}  a_i^2 =1$.
%\end{equation}\
 Our only requirement for the quantum channel is  that the two largest Schmidt coefficients are equal.
The state $\vert \Phi \rangle _{23}$ is a coherent superposition, with real  non-negative  probabilistic amplitudes, of a set of states,
\begin{equation}
\left\vert \Phi\right\rangle _{23}^{(\tau)}=
 \sum_{i=\tau}^{n}
 \frac{1}{\sqrt{n+1-\tau}}\left\vert i\right\rangle _{2} \left\vert i \right\rangle _{3},
\end{equation}
with $\tau=0,1,...,n-1$,
which are equivalent to $(n+1-\tau)$-dimensional maximally entangled states.
The set of  $\vert \Phi \rangle _{23}$ for a fixed $n$
is an $(n-1)$-dimensional polyhedron, with $n$ vertices corresponding to $\vert \Phi \rangle _{23}^{(\tau)}$, in the space of entanglement invariants  \cite{RevModPhys.81.865,Fei,RUI2011Alternative}.
Its entanglement entropy  \cite{Ent1996,Wootters98}  is lower bounded by 1,
and the lower bound is  attained by the state with  $a_{n-1}=a_{n}=1/\sqrt{2}$, or, say, the Bell state $\vert \Phi\rangle _{23}^{(n-1)}$.

The total state in the teleportation is given by
\begin{align}
\left\vert \Psi\right\rangle _{123}  & =\left\vert \phi\right\rangle_{1}\left\vert \Phi\right\rangle _{23} \nonumber \\&=
{\textstyle\sum\limits_{i=0}^{n}}(
a_{i}\left\vert 0i\right\rangle _{12}\alpha\left\vert i\right\rangle _{3}+
a_{i}\left\vert 1i\right\rangle _{12}\beta\left\vert i\right\rangle _{3}).
\end{align}
When $\vert \Phi\rangle _{23}=\vert \Phi\rangle _{23}^{(\tau)}$,  the basis of Alice's measurement can also be chosen according to the \textit{translation strategy} as  $2(n+1-\tau)$ entangled states
\begin{align}
&\left\vert \psi_{j\pm}^{(\tau)}\right\rangle _{12}=\frac{1}{\sqrt{2}}(\left\vert0j\right\rangle \pm\left\vert1, j+1  \right\rangle )_{12}, \nonumber\\
&\left\vert \psi_{n\pm}^{(\tau)}\right\rangle _{12}=\frac{1}{\sqrt{2}}(\left\vert0n\right\rangle \pm\left\vert1 \tau \right\rangle )_{12},
\end{align}
with $j=\tau,\tau+1,....,n-1$,
and $2\tau$ \textit{vanished}  product states  $\vert 0k\rangle$ and $\vert1k\rangle$, with $k=0,...,\tau-1$.
Just like the result of the two-qutrit channel in (\ref{psi3}), the basis for  the intermediate case between $\vert \Phi\rangle _{23}=\vert \Phi\rangle _{23}^{(\tau)}$ and $\vert \Phi\rangle _{23}=\vert \Phi\rangle _{23}^{(\tau+1)}$ can be derived by a unitary transformation acting on the subspace of $\{\vert 0n\rangle, \vert 1,\tau+1\rangle\}$.
Therefore, we surmise that the basis for a general case can be obtained
by a sequence of unitary operations in order as
\begin{align}\label{uk}
u_{k} &  =\left\vert\overline{0n}\right\rangle   \left\langle 0n\right\vert + \left\vert  \overline{1,k+ 1}\right\rangle  \left\langle 1,k+1\right\vert  \nonumber\\
&\ \ \  +(\openone -\left\vert 0n\right\rangle   \left\langle 0n\right\vert  -\left\vert 1,k+1\right\rangle  \left\langle 1,k+1\right\vert )
\end{align}
 on  $\vert \psi_{j\pm}^{(0)}\rangle _{12}$, with $k=0,...,n-2$ and  $j=0,...,n$.
 Here, the states $\vert\overline{0n} \rangle = c_{k} \vert 0n \rangle +s_{k} \vert 1,k+1 \rangle$  and $  \vert  \overline{1,k+ 1}  \rangle= c_{k} \vert 1,k+1 \rangle -s_{k} \vert 0n \rangle$, with
\begin{align} \textstyle{ c_k= \sqrt{\frac{1}{2}\left(1+\frac{a_{k}^{2}}{a_{k+1}^{2}}\right)}\ \ \  \mathrm{,} \ \ \ s_k  = \sqrt{\frac{1}{2}\left(1-\frac{a_{k}^{2}}{a_{k+1}^{2}}\right)}. \nonumber }\end{align}
That is, the bases are given by
 \begin{align}\label{psid}
 \!\!  \left\vert \psi_{j\pm}\right\rangle _{12}  & =u_{n-2}u_{n-3}\cdots u_{3}u_{2}u_{1}u_{0}\left\vert \psi_{j\pm}^{(0)}\right\rangle _{12} \nonumber\\
 &  = \left\vert 0j\right\rangle \pm c_{j}\left\vert1,j\!+\!1\right\rangle   \nonumber   \\
 &  \mp \! s_{j}\!  \left( \!
{\textstyle\prod\limits_{k=j+1}^{n-2}}
\! \!\!c_{k}\left\vert 0n\right\rangle  \!+\!\!
{\textstyle\sum\limits_{l=j+1}^{n-2}}
{\textstyle\prod\limits_{k=j+1}^{l-1}}
\! \!\!c_{k} s_{l}\left\vert 1,\! l\!+\!1\right\rangle \!\! \right),
%\\
 \end{align}
 \begin{align}
\left\vert \psi_{n\pm}\right\rangle _{12} &  =u_{n-2}u_{n-3}\cdots u_{0}\left\vert \psi_{n\pm}^{(0)}\right\rangle _{12}\nonumber\\
&  = \!  \left(
{\textstyle\prod\limits_{k=0}^{n-2}}
\!c_{k}\left\vert 0n\right\rangle   \!+\!
{\textstyle\sum\limits_{l=0}^{n-2}}
{\textstyle\prod\limits_{k=0}^{l-1}}
\!c_{k}s_{l}\left\vert 1,l\!+\!1\right\rangle \!  \right) \!\pm\! \left\vert 10\right\rangle,    \nonumber
\end{align}
 where $j=0,...,n-1$. Here, we omit
their normalization coefficients $(1/\sqrt{2})$ and the subscript identifying Alice's subsystems.

 One can prove that,  the teleportation can be accomplished perfectly by the joint measurement in the above orthonormal  basis (\ref{psid})
 by deriving the postmeasured states of qudit 3 left to Bob and showing they satisfy condition (ii).
 Namely, the (un-normalized) collapsed states are given by
\begin{align}
\left\vert \phi_{j\pm}\right\rangle _{3} &  = \alpha\biggr( a_{j}\left\vert j\right\rangle \mp s_{j}
{\textstyle\prod\limits_{k=j+1}^{n-2}}
c_{k} a_{n}\left\vert n\right\rangle \biggr)\nonumber\\
  \pm\beta& \biggr( \! c_{j} a_{j+1} \! \left\vert j\!+\!1\right\rangle\!-\!s_{\!j} \!
{\textstyle\sum\limits_{l=j+1}^{n-2}}
{\textstyle\prod\limits_{k=j+1}^{l-1}}
\! c_{k}s_{l}a_{l+1}\left\vert l\!+\!1\right\rangle \! \biggr), \nonumber\\
\left\vert \phi_{n\pm}\right\rangle _{3} &  =\alpha \biggr(
%TCIMACRO{\tprod \limits_{k=0}^{n-2}}
%BeginExpansion
{\textstyle\prod\limits_{k=0}^{n-2}}
%EndExpansion
c_{k}a_{n}\left\vert n\right\rangle \biggr) \\
&  +\beta \biggr(%
{\textstyle\sum\limits_{l=0}^{n-2}}
{\textstyle\prod\limits_{k=0}^{l-1}}
c_{k}s_{l}a_{l+1}\left\vert l+1\right\rangle \pm a_{0}\left\vert0\right\rangle\biggr),  \nonumber
\end{align}
where $j=0,...,n-1$.
And the  corresponding probabilities of Alice's outcome are
\begin{equation}\label{Pn}
P_{j\pm}=\frac{1}{2}(a_{j}^{2}+s_{j}^{2}
{\textstyle\prod\limits_{k=j+1}^{n-2}}
c_{k}^{2} a_{n}^{2}) \  \mathrm{,} \
P_{n\pm}=\frac{1}{2}
{\textstyle\prod\limits_{k=0}^{n-2}}
c_{k}^{2} a_{n}^{2}.
\end{equation}
A Direct calculation shows the  probability amplitudes in the above results satisfying
\begin{align}
 a_{j}^2 + s_{j}^2
{\textstyle\prod\limits_{k=j+1}^{n-2}}
c_{k}^2 a_{n}^2   &=
\! c_{j}^2 a_{j+1}^2+ \!s_{\!j}^2 \!
{\textstyle\sum\limits_{l=j+1}^{n-2}}
{\textstyle\prod\limits_{k=j+1}^{l-1}}
\! c_{k}^2s_{l}^2a_{l+1}^2, \nonumber\\
%%%
{\textstyle\prod\limits_{k=0}^{n-2}}
%EndExpansion
c_{k}^2 a_{n}^2
&=%
{\textstyle\sum\limits_{l=0}^{n-2}}
{\textstyle\prod\limits_{k=0}^{l-1}}
c_{k}^2s_{l}^2a_{l+1}^2  + a_{0}^2.  \nonumber
\end{align}
Therefore, they are of    the form $\alpha |\tilde{0}\rangle+\beta |\tilde{1}\rangle$.
Consequently, according to the classical information from Alice, Bob can transform these states to $|\phi\rangle_3$ perfectly using appropriate unitary operations, which are  independent of the state being teleported.

\section{Entanglement and classical information}	\label{entanglement}

%\textit{Entanglement and classical information.--}
The generation  and measurement of entangled states of high-dimensional  systems are  two technical challenges in laboratories \cite{2019Quantum,2020Experimental}.
 The process of sending  classical information to  Bob,
 the preparation of the quantum channel, and Alice's joint measurement can be regarded as three key ingredients consuming resources in the procedure.
 Our general scheme provides a continuous region to explore these resources in the perfect teleportation of a qubit, which consists of bipartite states whose two largest Schmidt coefficients are equal.

To measure these resources, we adopt the following three quantities analytically expressed in terms of  the Schmidt coefficients.
First, the entanglement entropy \cite{Ent1996} of the quantum channel is given by
\begin{equation}
\mathcal{E}\left(\left\vert \Phi\right\rangle _{23}\right)= - \sum_{i=0}^{n}  a^2_{i} \log_2 a^2_{i}.
\end{equation}
Second, the entanglement of Alice's joint measurement is defined by the average of the basis
\begin{equation}
\mathcal{E}_{12}=\sum_{j=0}^n \left[  P_{j +}\mathcal{E}\left(\left\vert \psi_{j+}\right\rangle _{23}\right)+ P_{j -}\mathcal{E}\left(\left\vert \psi_{j-}\right\rangle _{23}\right)  \right].
\end{equation}
It is a direct generalization of the definition in the work of  Li \textit{et al.} \cite{1999Probabilistic}, in which their four bases have the same entanglement degree.
The entanglement entropy of  the qubit-qudit states  (\ref{psid}) can be expressed as a monotone increasing function of their concurrences \cite{Wootters98},
 which are
\begin{align}
 &\mathcal{C}\left(\left\vert \psi_{j\pm}\right\rangle _{23}\right)=\sqrt{1- s_j^4 \prod_{k=j+1}^{n-2}c_k^4  },   \nonumber\\
& \mathcal{C}\left(\left\vert \psi_{n\pm}\right\rangle _{23}\right)=\prod_{k=0}^{n-2}c_k \sqrt{2-  \prod_{l=0}^{n-2}c_l^2  },
\end{align}
with $j=0,...,n-1$.
Third, the classical bits sent to Bob are given by  the Shannon entropy of the distribution (\ref{Pn}) as
\begin{equation}
\mathcal{H}_{12}=- \sum_{j=0}^n \left(  P_{j +} \log_2 P_{j +} +   P_{j -} \log_2 P_{j -}  \right).
\end{equation}
Below we list the results of  two series of one-parameter quantum channels, to show the properties of the three quantities more clearly.

\textit{Case I.}  The first $(n-1)$ Schmidt coefficients are equal.
Let  $x=a_0/a_{n} \in[0,1]$; one can find that only $u_{n-2}$ is nontrivial with the parameters $c_{n-2}=\sqrt{(1+x^2)/2}$ and
 $s_{n-2}=\sqrt{(1-x^2)/2}$, while the other unitary operations  $u_{k}=\openone$.
 Two pairs of the concurrence may be lower than $1$:
 \begin{align}
 \!\! \mathcal{C}\left(\left\vert \psi_{(n-2)\pm}\right\rangle _{23}\right)= \mathcal{C}\left(\left\vert \psi_{n\pm}\right\rangle _{23}\right)=\frac{1}{2}\sqrt{3+2x^2 -x^4}.
\end{align}
Their corresponding outcome probabilities are
 \begin{align}
P_{(n-2)\pm} =P_{n\pm} =\frac{1+x^2}{4[2+(n-1)x^2]} .
\end{align}
The other $(n-1)$ bases are equivalent to two-qubit Bell states with a concurrence of $1$,
whose outcome probabilities  are given by
 \begin{align}%\textstyle
 {
P_{k\pm}  =\frac{x^2}{2[2+(n-1)x^2]} \ \mathrm{,} \  P_{(n-1)\pm}  =\frac{1}{2[2+(n-1)x^2]}, \nonumber
}\end{align}
with $k=0,...,n-3$.

\textit{Case II.}  The first $(n-2)$ Schmidt coefficients are zero.
Since the results for $n=2$ are already incorporated into Case I,
we show only the nontrivial quantities for $n\geq 3$ here.
In addition, we adopt the convention that $c_k=1$ and $s_k=0$ when $a_k=a_{k+1}=0$.
Without special explanation, the unitary operations $u_{k}$ default to $\openone$, the values of the concurrence are $1$,
and the outcome probabilities are zero.
We set  $y=a_{n-2}/a_{n} \in[0,1]$.
The parameters in unitary operators $u_{n-3}$ and $u_{n-2}$ are $c_{n-3}=s_{n-3}=\sqrt{1/2}$, $c_{n-2}=\sqrt{(1+y^2)/2}$ and
 $s_{n-2}=\sqrt{(1-y^2)/2}$.
 There are three pairs of concurrences depending on the value of $y$,
 \begin{align}
 & \mathcal{C}\left(\left\vert \psi_{(n-3)\pm}\right\rangle _{23}\right)=   \sqrt{1-\frac{1}{16}(1+y^2)^2}, \nonumber\\
 & \mathcal{C}\left(\left\vert \psi_{(n-2)\pm}\right\rangle _{23}\right)=  \frac{1}{2}\sqrt{3+2y^2 -y^4}, \\
 & \mathcal{C}\left(\left\vert \psi_{(n-2)\pm}\right\rangle _{23}\right)=  \frac{1}{4}\sqrt{7+6y^2 -y^4},. \nonumber
\end{align}
 and four pairs of nonzero probabilities,
 \begin{align}%\textstyle
 {
&P_{(n-3)\pm}=P_{n\pm}    =\frac{1+y^2}{8(2+y^2)},\nonumber \\
&P_{(n-2)\pm}  =\frac{1+y^2}{4(2+y^2)}, \ P_{(n-1)\pm}  =\frac{1}{2(2+y^2)}.
}\end{align}

\begin{figure}
% \begin{flushright}
\includegraphics[width=8cm]{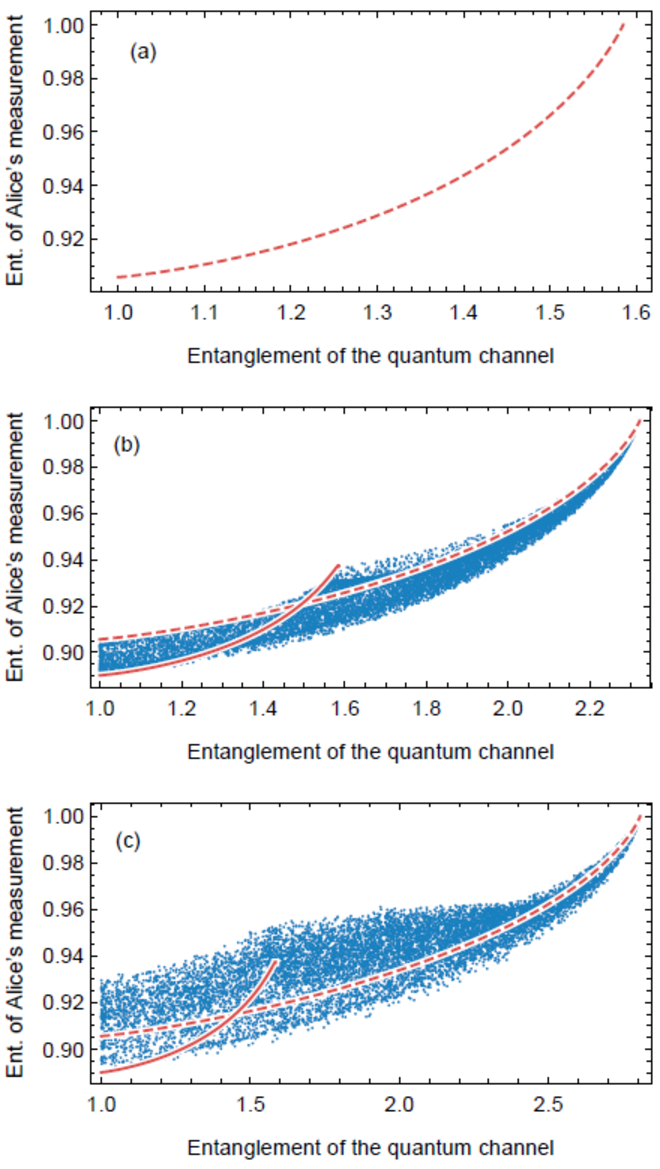}
%\end{flushright}
\caption{
Entanglement of Alice's measurement vs entanglement of quantum channels.
Dashed curves show the values for Case I, and solid ones show values for Case II,
accompanied by $10000$ random quantum channels with for each value of $n$:
(a) $n=2$, (b) $n=4$, and (c) $n=6$.
The results for $n=2$ overlap a  single line.
} \label{EE}
\end{figure}

\begin{figure}
% \begin{flushright}
\includegraphics[width=8cm]{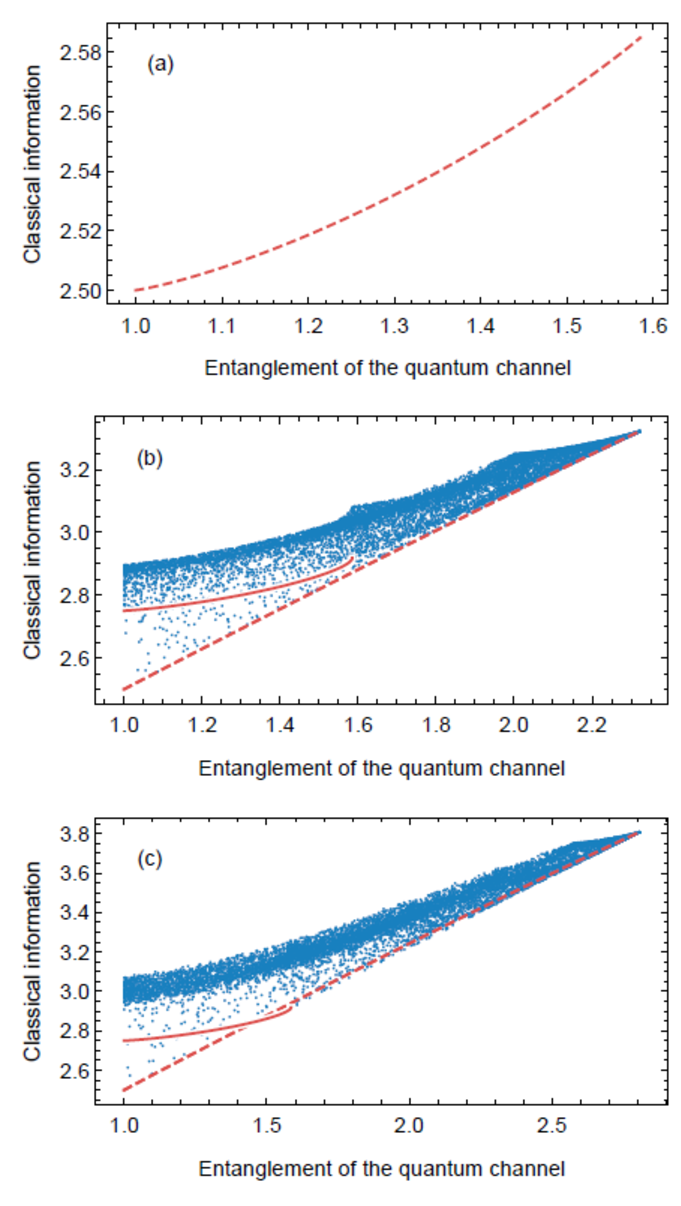}
%\end{flushright}
\caption{Classical bits sent to Bob vs entanglement of quantum channels with the same parameters as Fig.\ref{EE}.
} \label{BE}
\end{figure}

Figure \ref{EE} shows the regions of the entanglement degrees for Alice's measurement and the quantum channel, along with the curves for the two one-parameter cases, for a fixed dimension.
In general, the entanglement of the quantum channels is above the two-qubit Bell states with an entanglement entropy of $1$,
which is a necessary condition for perfect teleportation of a qubit according to Theorem $1$ in \cite{PRA2004},
while Alice's measurements are below them.
In this sense, the \textit{entanglement matching} \cite{1999Probabilistic}  does not appear in our scheme.
On the other hand, the entanglement of Alice's measurement has an overall upward trend  as the entanglement of a quantum channel increases for a fixed $n$.
It reaches the maximum of $1$, when the channel is maximally entangled.
Hence, the entanglement of Alice's measurement can be considered Alice's ability to use the quantum channel,
and a stronger ability is required for a more entangled quantum channel.

 For a fixed $n$, the minimal entanglement of Alice's measurement occurs at the limit of  Case II  with $y\rightarrow 0$ and $a_{n-1}=a_{n}\rightarrow\sqrt{1/2}$.
 The minimums can be derived directly from the above analytic expressions as
 $\mathcal{E}_{12}=\frac{1}{2} H\left(\frac{3}{4}\right)+\frac{1}{2}\approx0.906$ when $n=2$ and
   $\mathcal{E}_{12}=\frac{1}{4} H\left(\frac{3}{4}\right)+\frac{1}{4} H\left(\frac{15}{16}\right)+\frac{1}{2}\approx0.890$  when $n=3,4,...$,
where $H(t) = -\frac{1}{2} \left(1-\sqrt{1-t}\right) \log _2 \frac{1}{2} \left(1-\sqrt{1-t}\right) -\frac{1}{2} \left(1+ \sqrt{1-t}\right) \log _2 \frac{1}{2} \left(1+ \sqrt{1-t}\right)$.
Although the minimum is  independent of the dimension of the channel when  $n\geq3$,
the minimal entanglement in Alice's single measurement can be found to decrease with $n$.
Namely, it is the entanglement of $\left\vert \psi_{n\pm}\right\rangle _{23}$ in the limit of  $a_{k}/a_{k+1}\rightarrow0$ and $a_{k}\rightarrow0$ and
equals  $H[2^{-(n-3)}- 2^{-(2n-4) }]$.
This indicates that, for a large $n$, one can design a protocol to detect such a small amount of entanglement to teleport a qubit exactly.
However, the successful probability also decreases quickly with $n$ as $P_{n+} +P_{n-}    =2^{-(n-1)}$.

One can also notice that the entanglement of Alice's measurement corresponding to  a quantum channel $\left\vert \Phi\right\rangle _{23}^{(\tau)}$ is smaller than that of an $(n+1-\tau)$-dimensional maximally entangled state.
The reason can be found in the example of two-qutrit channels, where we superpose $\vert \psi^{(b)}_{2\pm} \rangle_{12} $ with the two \textit{vanished} vectors, $\left\vert00\right\rangle $ and $\left\vert10\right\rangle $.
This reduces the entanglement of Alice's measurement while two outcomes are added, which increases the classical information sent to Bob.
That is, the price to decrease the entanglement of Alice's measurement is to send more classical bits to Bob.
This conclusion can be confirmed by comparing Fig.\ref{EE} and \ref{BE}.

 Obviously, the classical information also has an overall upward trend with the entanglement of the quantum channel.
The minimum  of classical bits  can be found  at the limit of  Case I  with $x\rightarrow 0$, which is $5/2$ and independent of $n$.

For a fixed $n$, at the left end points of the curves for the two cases, which  correspond to the same quantum channel but different measurements,
Alice  measures more entanglement in Case I  than in Case II, but  she sends more classical bits to Bob in Case II.
In addition, the classical bits in Case I increase faster than in Case II, while for the behaviors of entanglement the opposite is true.
In conclusion, the classical information sent to Bob  is also a necessary resource to use  the quantum channel,
which appears to be complementary  to the entanglement of Alice's measurement.

 \section{Discussion}\label{disscuss}

We provide some qualitative discussion of the above results in this part.
For simplicity,  we show only the formulas for the two-qutrit channel, although they can be directly extended to the general case.

One can perform the unitary operation $u_0$ with $n=2$ in (\ref{uk}) on the maximally entangled  two-qutrit  states \cite{2019Quantum}
 and obtain a set of bases as
 \begin{align}\label{psi33}
\left\vert \psi_{0j}\right\rangle _{12} &  =\frac{1}{\sqrt{3}}[\left\vert00\right\rangle  +\omega^j  (c\left\vert 11\right\rangle -s\left\vert 02\right\rangle) + \omega^{2j} \left\vert 22\right\rangle]_{12}, \ \ \ \ \ \  \ \ \  \nonumber\\
\left\vert \psi_{1j}\right\rangle _{12} &  =\frac{1}{\sqrt{3}}[\left\vert 10 \right\rangle +\omega^j \left\vert 21\right\rangle +\omega^{2j}  (c\left\vert 02\right\rangle + s\left\vert 11\right\rangle)]_{12}, \ \ \  \ \ \  \ \ \  \nonumber \\
\left\vert \psi_{2j}\right\rangle _{12} &  =\frac{1}{\sqrt{3}}[\left\vert 21 \right\rangle + \omega^j  \left\vert 01\right\rangle  + \omega^{2j} \left\vert 12\right\rangle ]_{12},
\end{align}
 where $\omega=\exp (i\frac{2\pi}{3})$, $j=0,1,2$, and $c$ and $s$ are defined in (\ref{cs}).
Measurement on such a basis can teleport a qutrit state
\begin{equation}
\left\vert \phi\right\rangle _{1}^{(3)}=\alpha\left\vert 0\right\rangle _{1}+\beta\left\vert 1\right\rangle _{1}+ \gamma \left\vert2\right\rangle _{1},
\end{equation}
with a state-dependent fidelity,
via the two-qutrit  channel studied in Sec. \ref{qutrit}.
Namely, the measurement collapses the three-qutrit state  $\left\vert \phi\right\rangle _{1}^{(3)} \left\vert \Phi\right\rangle _{23}$ into
\begin{align}
\left\vert \phi_{0j}\right\rangle _{3} &  \propto [\alpha(a_{0}\left\vert 0\right\rangle - \omega^j a_{1}s\left\vert
2\right\rangle ) +  \beta a_{1}c \omega^j \left\vert 1\right\rangle    +    \gamma a_1 \omega^{2j} \left\vert 2\right\rangle  ]_{3},\nonumber\\
\left\vert \phi_{1j}\right\rangle _{3} &  \propto[ \alpha c  a_{1}\omega^{2j} \left\vert 2\right\rangle
+ \beta (a_{0}\left\vert 0\right\rangle +  a_{1}s \omega^{2j} \left\vert1\right\rangle )
+\gamma a_1  \omega^j  \left\vert1\right\rangle
] _{3} , \nonumber\\
\left\vert \phi_{2j }\right\rangle _{3}&  \propto[ \gamma \alpha a_{0} \left\vert 0\right\rangle + \alpha a_{1} \omega^j \left\vert 1\right\rangle + \beta a_{1} \omega^{2j} \left\vert 2\right\rangle] _{3}. \nonumber
\end{align}
In each collapsed state, the vectors multiplied by the coefficients  $\alpha$ and $\beta$ are orthogonal and equal in magnitude.
When $\gamma=0$, according to Alice's outcome,
Bob can rebuild the state $\left\vert \phi\right\rangle _{1}^{(3)}$ with a fidelity of $1$  by performing appropriate unitary operations on qutrit $3$.
These unitary operations are similar to those in Sec. \ref{qutrit}, with being replaced by $\omega^{0,1,2}$.
However, when $\gamma\neq0$, the same operations bring him only a state with a fidelity less than $1$,  which is dependent on the initial state, quantum channel, and Alice's outcome.
 We derive the average fidelity over Alice's outcomes and the initial states under the Haar measure \cite{randomU} as
\begin{equation}
\left\langle \mathcal{F} \right\rangle = \frac{7}{3}+\frac{5}{2}a_1^2 +a_0 a_1 -\frac{5}{3(1-a_1^2)}.
\end{equation}
It increases from $1/4$ to $1$ as  $a_0$ increases from $0$ to $1/\sqrt{3}$.

The perfect  protocol studied in Sec. \ref{qutrit} can be regarded as an improved version of the above imperfect teleportation
 with the aid of \emph{a priori} knowledge of $\gamma=0$.
The prior knowledge reduces the entanglement of Alice's measurement and classical bits;
for example, the nine maximally entangled two-qutrit bases are  replaced by the six maximally entangled  states (\ref{psia}) in
subspaces when $a_0=1/\sqrt{3}$.

The behaviors of  Alice's capabilities in the perfect  protocol can be understood by using the relation corresponding to the imperfect teleportation.
Taking  the standard teleportation of a qutrit \cite{2019Quantum} as a reference,
the unitary operator $u_0$, creating the basis (\ref{psi33}),
 from the maximally entangled states
concentrates the fidelity into the subspace of $(\alpha, \beta)$
and simultaneously decreases the entanglement of measurement and classical information.
Similarly, the same unitary operator also decreases the two quantities in the perfect protocol  as it transforms
$ \vert \psi^{(a)}_{i\pm} \rangle_{12} $ in (\ref{psia}) into $ \vert \psi_{0\pm} \rangle _{12} $  in (\ref{psi3}).
This leads to a smaller entanglement of measurement than the Bell-state measurement,
while the classical bits are greater than $2$ as the price of increasing the dimension.
The increasing of  the entanglement of $\vert \Phi \rangle _{23}$ enhances its ability to teleport the state $ \vert \phi \rangle _{1}^{(3)}$;
therefore the unitary operator $u_0$ which ensures the fidelity in the subspace of $(\alpha, \beta)$ becomes  \emph{smaller}.
Correspondingly, this increases the entanglement of Alice's measurement and  classical bits  in the perfect protocol.

\begin{figure}
% \begin{flushright}
\includegraphics[width=8cm]{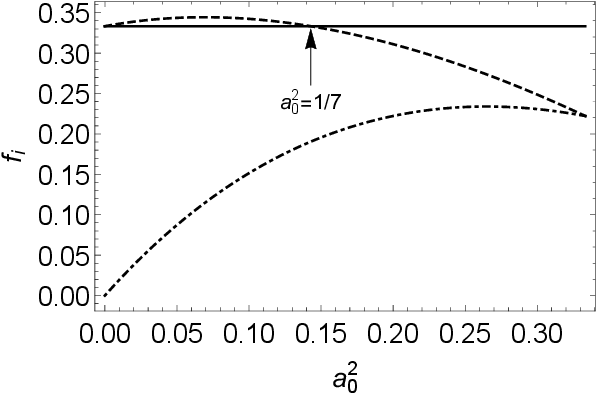}
 %\end{flushright}
\caption{
%The amounts of $f_i$
The responses of fidelities to the random phases vs $a_0^2$.
The dashed and dot-dashed curves show $f_0$  and $f_1$ in (\ref{fi}), respectively, and the solid line is for the value of $1/3$.
} \label{fidelity}
\end{figure}

It is also necessary  to compare our protocol in Secs. \ref{qutrit} and \ref{qudit}
and the standard teleportation of a qubit via a Bell state under the influence of environmental noise.
We now show our protocol is more robust than the standard one under some specific noise channels.
Namely,
we assume that qutrit $3$ passed through an inhomogeneous phase noise when it was sent to Bob,
which brings each ket a random phase as $|j\rangle \rightarrow e^{i \theta_j} |j\rangle$, with $j=0,1,2$.
Suppose $\langle  e^{i \theta_j}  \rangle = 1-q_j$ and $q_j\in [0,1]$.
One can derive the fidelity of teleportation under the noise
 by calculating  the overlaps between Bob's collapsed states and their ideal forms in (\ref{psiipm}),
and then its average value $\left\langle \mathcal{F} \right\rangle $ over the initial states.
When $q_{j'\neq j} =0$, the average fidelity can be expressed as
$
\left\langle \mathcal{F} \right\rangle = 1 -  (1-q_j) f_j
$,
and
\begin{eqnarray}\label{fi}
f_0= \frac{1+2 a_0^2 - 7 a_0^4}{3(1+a_0^2)}, \ \ \
f_1=f_2= \frac{ 2 a_0^2 ( 3  - 5a_0^2 )}{3(1+a_0^2)}.   \ \ \
\end{eqnarray}
In the same one-sided channel,
the fidelity of the standard scheme has a similar form with $f_0=f_1=1/3$ and $f_2=0$.
Here, $f_2=0$ is due simply to the absence of $|2\rangle$ in the standard scheme.
These amounts of $f_i$ quantify the responses of fidelities to the random phases,
which are shown in Fig. \ref{fidelity}.
When the random phase occurs in $|1\rangle$, our two-qutrit protocol is more robust than the standard teleportation via a Bell state.
When it occurs in $|0\rangle$, the former performs better than the latter when $a_0^2>1/7$,
while the latter is more robust when $a_0^2<1/7$.

 \section{Summary}\label{summ}

%\textit{Summary.--}
We presented a scheme for perfect teleportation of a qubit by using a two-qudit pure state with in which the two Schmidt coefficients are equal.
For a fixed dimension $d$, the quantum channels, no longer confined to the maximally entangled states, form a continuous area in the space of entanglement invariants.
To play the role of a quantum channel, the entangled state requires an appropriate joint measurement by Alice and enough classical bits being sent to Bob.
Our scheme requires less entanglement of Alice's measurement but more classical bits than the standard teleportation via a Bell state.
The two quantities increase with the entanglement of the quantum channel
and thereby can be regard as Alice's necessary capabilities to use the  channel.
The two capabilities appears complementary to each other.
For a fixed amount of entanglement of the quantum channel, the entanglement in Alice's measurement can be partially replaced by classical bits sent to Bob.
We also provided an intuitive understanding of these behaviors by using imperfect teleportation of a qutrit,
in which the fidelity in a subspace is  ensured to be $1$.
Under some specific noise channels, our protocol is more robust than the standard teleportation.

It would be interesting to consider the following open questions or extensions.
First, can our scheme be generalized to the teleportation of high-dimensional states?
We conjecture that a $d^{\prime}$-level state can be teleported perfectly by using a two-qudit ($d\geq d^{\prime} $)  pure state in which the $d^{\prime}$ largest Schmidt coefficients are equal.
However, we find that the joint measurement for high-dimensional teleportation may not be constructed by a simple generalization of (\ref{psid}),
and therefore, the development of  new methods to design the protocol is still necessary.
Second,  while we focused here on the teleportation between two participants, hybridizing the present ideas to controlled teleportation would  be interesting.
Third,  implementing  the protocol experimentally is a natural direction.
Besides the generation of  high-dimensional quantum channels, both Alice's measurement and Bob's operations are challenges in the laboratory.
The techniques developed in recent experiments \cite{2019Quantum,2020Experimental} open the possibility of implementating the process in an optical system.

 \begin{acknowledgments}
 %\textit{Acknowledgments.--}
This work was supported by the National Natural Science Foundation of China (Grants No. 11675119, No. 11575125, and No. 11105097) and  the Fundamental Research Funds for the Central Universities (Grant No. 2020JKF306).
 \end{acknowledgments}

  \bibliography{QunitQubit}

\end{document}